\documentclass[aps,prb]{revtex4}
\usepackage{amsmath,amssymb}
\usepackage{graphicx}
\usepackage{dcolumn}
\usepackage{bm}
\usepackage{color}
\newcommand{\mbb}{\mathbb}
\newcommand{\mc}{\mathcal}

\newcommand{\tet}{\texttt}
\newcommand{\pr}{\partial}

\begin{document}
\title{Quantum-statistical theory for laser-tuned transport and optical conductivities of dressed electrons in $\alpha-\mc{T}_3$ materials}

\author{Andrii Iurov$^{1,2} \footnote{E-mail contact: aiurov@unm.edu, theorist.physics@gmail.com
}$,
Liubov Zhemchuzhna$^3$,
Dipendra Dahal$^3$,
Godfrey Gumbs$^{3,4}$, 
and
Danhong Huang$^{5,2}$
}

\affiliation{
$^{1}$Department of Physics and Computer Science, Medgar Evers College of City University 
of New York, Brooklyn, NY 11225, USA \\
$^{2}$Center for High Technology Materials, University of New Mexico,
1313 Goddard SE, Albuquerque, New Mexico, 87106, USA \\
$^{3}$Department of Physics and Astronomy, Hunter College of the City
University of New York, 695 Park Avenue, New York, New York 10065, USA\\
$^{4}$Donostia International Physics Center (DIPC),
P de Manuel Lardizabal, 4, 20018 San Sebastian, Basque Country, Spain\\
$^{5}$Air Force Research Laboratory, Space Vehicles Directorate,
Kirtland Air Force Base, New Mexico 87117, USA}

\date{\today}

\begin{abstract}
In the presence of external off-resonance and circularly-polarized irradiation, we have derived a many-body formalism and performed a detailed numerical analysis for 
both the conduction and optical currents in $\alpha-\mc{T}_3$ lattices. The calculated complex many-body dielectric function, as well as conductivities of 
displacement and transport currents, display strong dependence on the lattice-structure parameter $\alpha$, especially approaching the graphene limit with $\alpha \to 0$.
Unique features in dispersion and damping of plasmon modes are observed with different $\alpha$ values, which are further accompanied by a reduced transport conductivity under irradiation. 
The discovery in this paper can be used for designing novel multi-functional nanoelectronic and nanoplasmonic devices.  
\end{abstract}

\maketitle

\section{Introduction} 
\label{sec-1}

So far, the $\alpha-\mc{T}_3$ model seems to present prospective opportunities for revolutionizing low-dimensional physics through novel 
two-dimensional (2D) materials.\,\cite{piech1} Its atomic configuration consists of a graphene-type honeycomb lattice along with an additional
site, i.e., a hub atom at the center of each hexagon.\cite{vidal} An essential structure parameter $\alpha = \tan \phi$, which enters into the
low-energy Dirac-Weyl pseudospin-1 Hamiltonian for $\alpha-\mc{T}_3$ model, is found to be the ratio between the rim-to-hub and rim-to-rim hopping coefficients. This parameter
affects all fundamental electronic properties of the $\alpha-\mc{T}_3$ lattice through topological characteristics embedded in its pseudospin-1 wave functions. 
Parameter $\alpha$ can vary from $0$ to $1$, corresponding to different types of $\alpha-\mc{T}_3$ 
materials, and the control of it could lead to some important technological applications for electronic and optoelectronic devices. Here, the case with $\alpha=0$ relates to graphene 
with a completely separated flat band, whereas $\alpha=1$ results in a pseudospin-1 dice lattice which has been fabricated and studied 
considerably.\,\cite{dice2,dice1} Consequently, the $\alpha-\mc{T}_3$ model may be viewed as an interpolation between graphene and the dice 
lattice (or pseudospin-1 $\mc{T}_3$ model).  Its low-energy dispersion consists of a Dirac cone, similar to that for graphene,\,\cite{gr01}
as well as a flat band with zero-energy separating the valence from the conduction band for these pseudospin-1 materials.\,\cite{flat1, malcolmMain}

\medskip
\par 
In recent years,  there have been numerous attempts for experimental realization of the $\alpha-\mc{T}_3$ model.  Its topological
characteristics, i.e., a Dirac cone with three bands touching at a single point, was observed in the triplon band structure of SrCu$_2$(BO$_3$)$_2$,
as an example of general Mott-Hubbard insulators.\,\cite{Dan5} Moreover, dielectric photonic crystals with zero refractive index also display Dirac cone dispersion at 
the center of the Brillouin zone under an accidental degeneracy.\,\cite{Dan6,Dan7} Most importantly, there exist various types of 
photonic Lieb lattices,\,\cite{Dan1,Dan2} consisting of a 2D array of optical waveguides. Such waveguide-lattice structure is shown to have a three-band structure, 
including a perfectly flat middle band.

\medskip
\par 
Further to a relatively recent proposal on $\alpha-\mc{T}_3$ model, there have been a lot of crucial publications devoted to investigating their magnetic,\,\cite{piech1, thesis, nic1, nic2} optical,\,\cite{dey1, dey2} 
many-body\,\cite{malcolmMain} and electron transport properties,\,\cite{tutul1,tutul2,Dutta1,Dora1} as well as to generalized versions of this model\,\cite{piech2}. 
A number of compelling properties of graphene and other low-dimensional
materials\,\cite{SilMain} have been realized in $\alpha-\mc{T}_3$ materials, including Klein tunneling\,\cite{kats, nic3} and Hofstadter
butterfly\,\cite{ourB}. Meanwhile, all pseudospin-1 structures also display some previously unknown phenomena resulting from the existence of a flat band 
in their energy dispersions\,\cite{gus19}, e.g., distinctive plasmon modes with a branch ``{\em pinching\/}'' feature at the Fermi level.\,\cite{malcolmMain} 

\medskip
\par 
We note that an exciting emergent technical application for condensed-matter quantum optics is {\it Floquet engineering}. This subject leads to a wide-range
optical-tuning capability and control of electron optical and transport currents in 2D materials by introducing an off-resonant periodic dressing field
in either terahertz or microwave frequency.\,\cite{p1,p2,p3,p4,Tor1,Tor2,ior1} Such external irradiation imposed on a 2D material produces a dramatic change
in most of its electronic properties due to creating so-called {\it dressed states}. Thisgives rise to a single quantum entity, consisting of an electron 
interacting  with a photon. It is described by unique energy-dispersion relations, depending on the intensity and polarization of incoming radiation.

\medskip
\par 
Our investigation of these electron dressed states is based on Floquet theory for quantum systems, driven by external periodic potentials. This
results in a $\backsim 1/(\hbar \omega)$ series expansion, as employed by Floquet-Magnus, Brillouin-Wigner and others,\,\cite{prX1} and provides
an effective analytical tool for investigating light-electron interaction in a variety of novel 2D materials\,\cite{kiMain,oura,ki1210,ki-spri} 
and optically-induced topological surface states\,\cite{ki19R} as well.

\medskip
\par
The modifications of single-particle band structure and wave function greatly affect the many-body dielectric function\,\cite{ourcontrolling, SilMain, ourPQE} 
in addition to electron conduction current\,\cite{Mor1, Mor2} and conductivity\,\cite{kisrep}. 
These changes mainly come from opening an energy gap\,\cite{kiMain} between the valence and conduction bands,
as well as from topologically-modified wave functions.\,\cite{pavlo} For a linearly-polarized optical dressing field, we find strong 
in-plane anisotropy, and even the anisotropic dispersion of phosphorene still experiences a lot of changes under this dressing field.\,\cite{ourJAP2017}  

\medskip
\par 
Once Floquet engineering has been applied to $\alpha-\mc{T}_3$ materials, we expect some fundamental changes will occur, such as, band-dispersion anisotropy, 
opening inequivalent energy gaps within each band, breaking down the electron-hole symmetry and valley degeneracy,\,\cite{dey1} and topological
variation of electron wave functions including their symmetries and the Berry phases.\,\cite{ourpeculiar} Interestingly,  circularly-polarized
radiation can induce a topological phase transition from a gapless semimetal to a topological insulator with a nonzero Chern number.\,\cite{dey2} This result
acquires a resemblance to a topological insulator, obtained from a periodic array of quantum rings under a circularly-polarized optical 
field.\,\cite{ki18top}
\medskip
\par 

The rest of this paper is organized as follows. In Sec.\,\ref{sec-2}, electron dressed states by an strong laser field in $\alpha-\mc{T}_3$ materials are derived.
In Sec.\,\ref{sec-3}, the effects of optically-dressed states on many-body dielectric function, plasmon mode and electron optical current are presented. 
The dressed-state effects on rate for elastic impurity scattering and conduction current are displayed in Sec.\,\ref{sec-4}.
Finally, a summary and some remarks are given in Sec.\,\ref{sec-5}.

\section{Valley-Spin dependent dressed states}
\label{sec-2}

The optical current of electrons in $\alpha-\mc{T}_3$ materials should be driven by a laser field.
In the presence of the laser field, bare electron states are obtained by substituting the electron wavevector $\mbox{\boldmath{$k$}}=\{k_x,k_y\}$ in the Hamiltonian of considered materials with 
$\mbox{\boldmath{$k$}}-(e/\hbar)\,\mbox{\boldmath{$A$}}(t)$. 
Here, $\mbox{\boldmath{$A$}}(t) = (\mc{E}_0/\omega)\left\{\cos (\omega t), \sin (\omega t)\right\}$ is a spatially-uniform vector potential associated with the applied circularly-polarized light,
where $\mc{E}_0$ is the amplitude of the electric-field component of imposed radiation and $\omega$ is its angular frequency in the off-resonance regime. 
As a result, the Hamiltonian will acquire an additional time-dependent term due to light-electron interaction. 

\medskip
\par
In this paper, we employ the Floquet-Magnus perturbation-expansion theory for our calculations. This procedure is applicable to any periodically-driven quantum 
structure\,\cite{prX1} so as to obtain valley-degenerate, isotropic and symmetric energy bands near the flat band,\,\cite{ourpeculiar} yielding 
$\varepsilon_0(k,\lambda_0) = 0$ and  

\begin{figure} 
\centering
\includegraphics[width=0.6\textwidth]{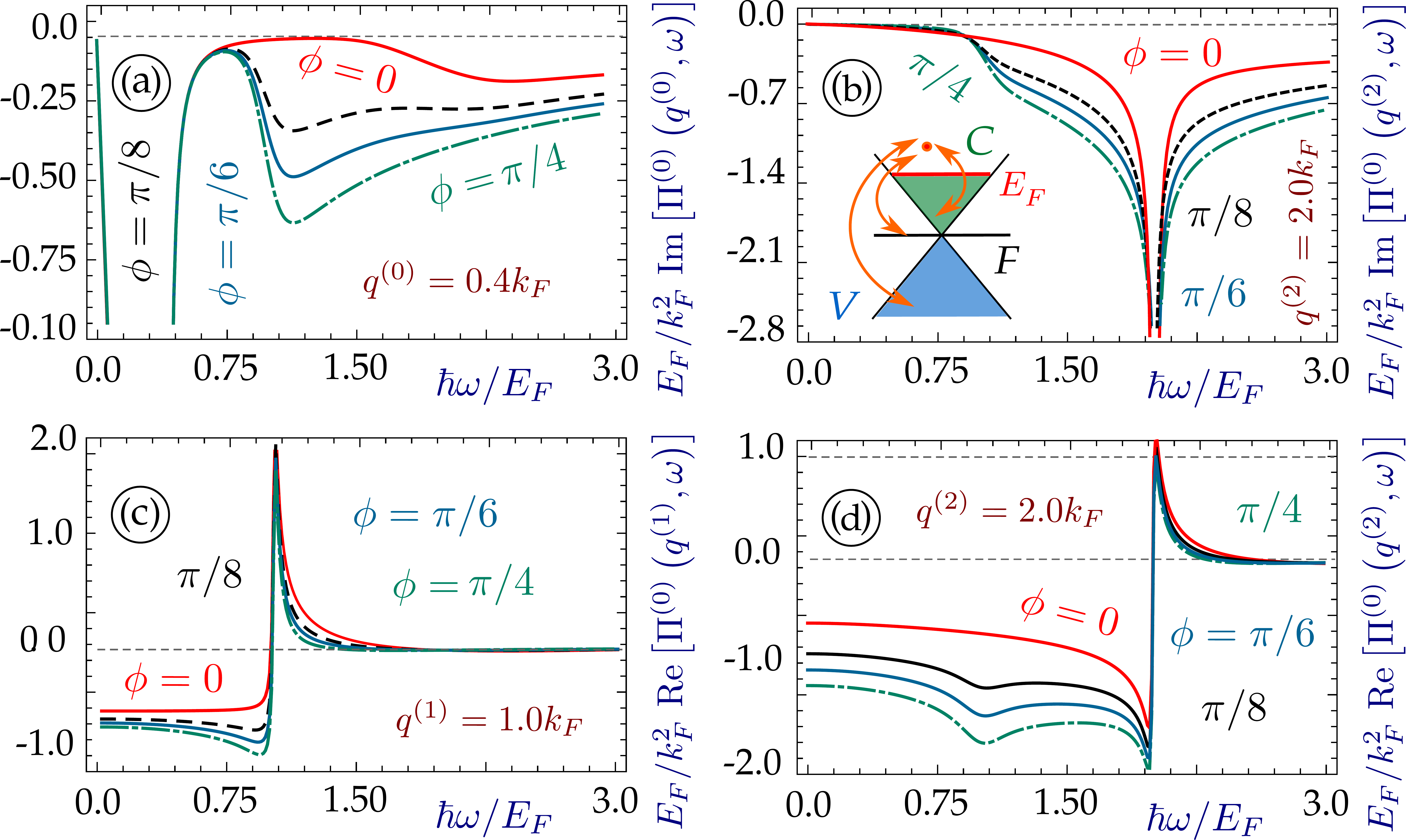}
\caption{(Color online) Frequency ($\omega$) and wavevector ($q$) dependent polarization function $\Pi_0(q, \omega \, \vert \, \phi)$ (in units of $k_F^2/E_F$) for non-irradiated $\alpha-\mc{T}_3$ model ($\lambda_0=0$) as a function of $\hbar\omega$
with various phase values $\phi$. Here, each panel corresponds to a chosen wavevector $q=q_i$ and each curve is for a specific phase $\phi$, as labeled. 
Two upper panels [$(a)$, $(b)$] present the imaginary part of $\Pi_0(q, \omega \, \vert \, \phi)$, whereas the two lower ones [$(c)$, $(d)$] are its real part. 
The inset in plot $(b)$ illustrates all possible single-particle transitions (red double-arrow curves) contributing to $\Pi_0(q, \omega \, \vert \, \phi)$ at $T=0\,$K with a horizontal red line for the Fermi energy $E_F$.}
\label{FIG:1}
\end{figure}

\begin{equation}
\label{zero}
\varepsilon_\gamma(k,\lambda_0) = \gamma\sqrt{\left(\lambda_0 c_0 / 2 \right)^2 + \left[ \hbar v_F k \, (1 -
\lambda_0^2 / 4) \right]^2 \,}\ ,
\end{equation}
where $\gamma=\pm$ for electrons ($+$) and holes ($-$), $c_0 = e\mc{E}_0v_F /\omega$ represents the interaction coefficient (energy), $v_F$ is the Fermi velocity, 
and $\lambda_0 = c_0/(\hbar \omega) = e\mc{E}_0v_F/(\hbar \omega^2)$ 
is a dimensionless light-electron coupling parameter. We limit our consideration to off-resonance frequencies of the laser field,
where the photon energy $\hbar\omega$ greatly exceeds any electron energies, e.g., Fermi energy $E_F=\hbar v_Fk_F$ with the Fermi wavenumber $k_F=\sqrt{\pi\rho_0}$ 
and areal doping density $\rho_0$. Consequently, we have $\lambda_0 \ll 1$
in spite of the light intensity $\mbb{I}_0 = \epsilon_0 c\,\mc{E}_0^2/2 \backsim 10 \, W/cm^2$.

\medskip
\par
The obtained dispersions in Eq.\,\eqref{zero} reveal an energy bandgap $E_G=\lambda_0 c_0\equiv 2\Delta_0$ which is only half of the 
graphene gap energy with the same interaction coefficient $c_0$.\,\cite{kiMain} The electronic states for an irradiated dice lattice, 
pertaining to the valence and conduction bands, are given by

\begin{equation}
\label{Twf1}
\Psi^\tau_\gamma (\mbox{\boldmath{$k$}},\lambda_0) = \frac{1}{\sqrt{\mc{N}^\tau_{\gamma}}} \, \left[
 \begin{array}{c}
  \tau \, \mc{C}^\tau_{1,\gamma} \, \tet{e}^{ - i \tau \theta_{\bf k}} \\
  \mc{C}^\tau_{2,\gamma} \\
  \tau \,(\hbar v_F k)^2 \, \tet{e}^{ + i \tau \theta_{\bf k}} 
 \end{array}
 \right] \, ,
\end{equation}
where

\begin{eqnarray}
 &&  \mc{C}^\tau_{1,\gamma}(k,\lambda_0) = (\hbar v_F k)^2 + 2 \left(
 \delta_{\lambda}^2 -  \gamma\,\tau\delta_{\lambda}\sqrt{
 (\hbar v_F k)^2 + \delta_{\lambda}^2}\right)\ , \\
 \nonumber 
 &&  \mc{C}^\tau_{2,\gamma}(k,\lambda_0) = \sqrt{2} \, \gamma \, (\hbar v_F k) \, \left(
 \sqrt{
 (\hbar v_F k)^2 + \delta_{\lambda}^2
 } - \gamma \, \tau\delta_{\lambda}
 \right) \ , \\
 \nonumber 
 && \mc{N}^\tau_{\gamma}(k,\lambda_0 \ll 1) \backsimeq 
  4 \,(\hbar v_F k)^4 - 4 \gamma\,\tau \, c_0 \lambda_0 \, (\hbar v_F k)^3 + 3 \left[
  c_0 \lambda_0 \,(\hbar v_F k) \right]^2 +\cdots\ \ .
 \end{eqnarray}
Here, the parameter $\delta_\lambda = 2 \lambda_0 c_0 / (4 - \lambda_0^2)$ is different from the actual energy gap 
$E_{G} =\lambda_0 c_0=2\Delta_0$. 
For the flat band with $\gamma = 0$, on the other hand, three components of its wave function are not the same and the middle one is nonzero, as
expected for a finite energy gap (see Appendix\ \ref{app-1}).

\begin{figure}
\centering
\includegraphics[width=0.6\textwidth]{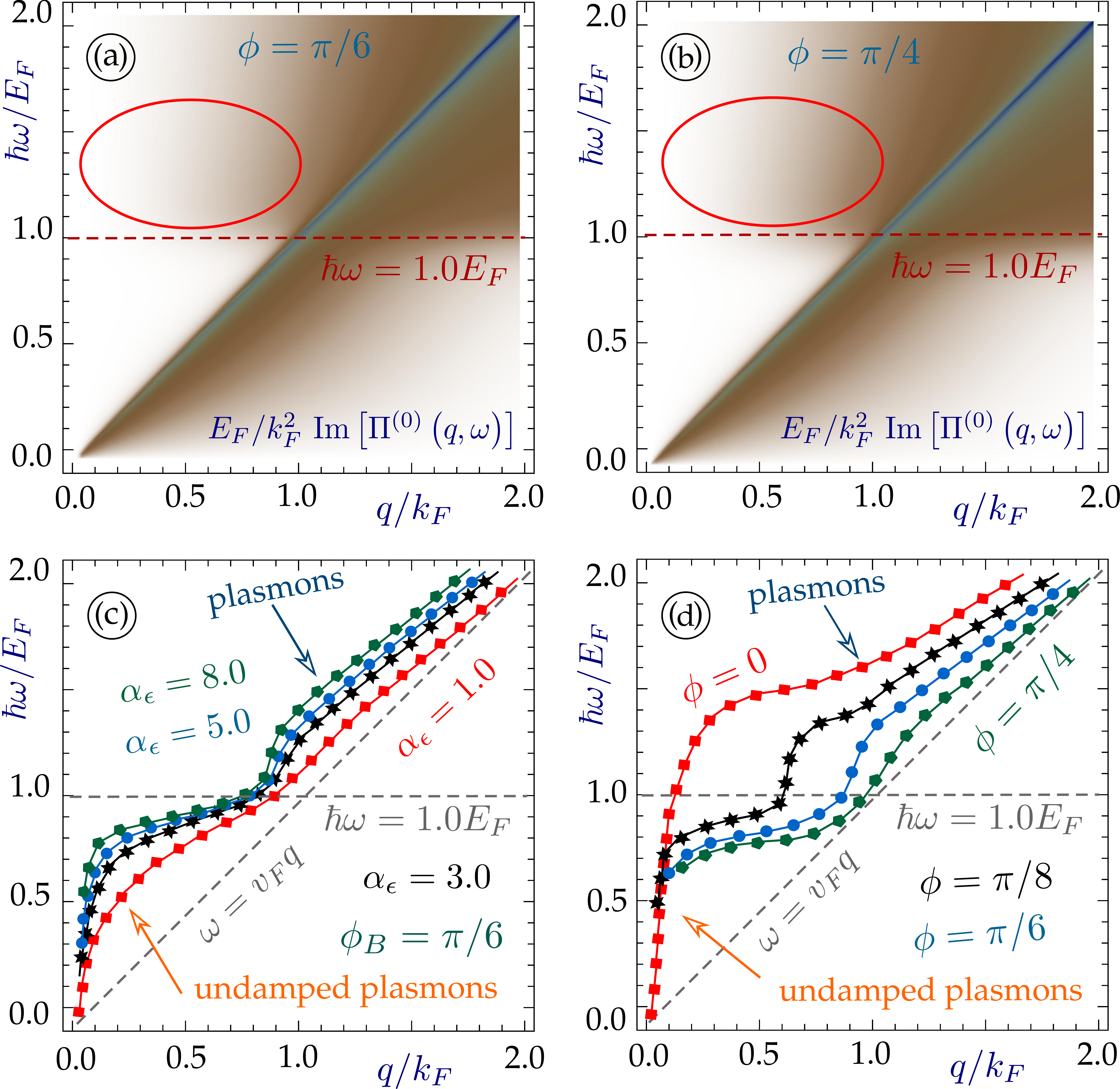}
\caption{(Color online) Plasmon damping regions [$(a)$ for $\phi=\pi/6$, $(b)$ for $\phi=\pi/4$] and plasmon branches [$(c)$, $(d)$] for various types of 
$\alpha-\mc{T}_3$ lattices in the absence of external irradiation ($\lambda_0=0$). Two upper panels display $\text{Im}[\Pi_0(q, \omega \, \vert \, \phi)]$, 
which indicates regions for single-particle excitations. Here, two partially-damping regions above the $\hbar\omega=E_F$ line are highlighted by red circles. 
Panel $(c)$ presents plasmon dispersions for fixed $\phi = \pi/6$ but different $\alpha_0$ values, while panel $(d)$ 
exhibits plasmon modes for fixed  $\alpha_0 = 3.0$ and several phase values of $\phi$.
}
\label{FIG:2}
\end{figure}  
 
\begin{figure}
\centering
\includegraphics[width=0.6\textwidth]{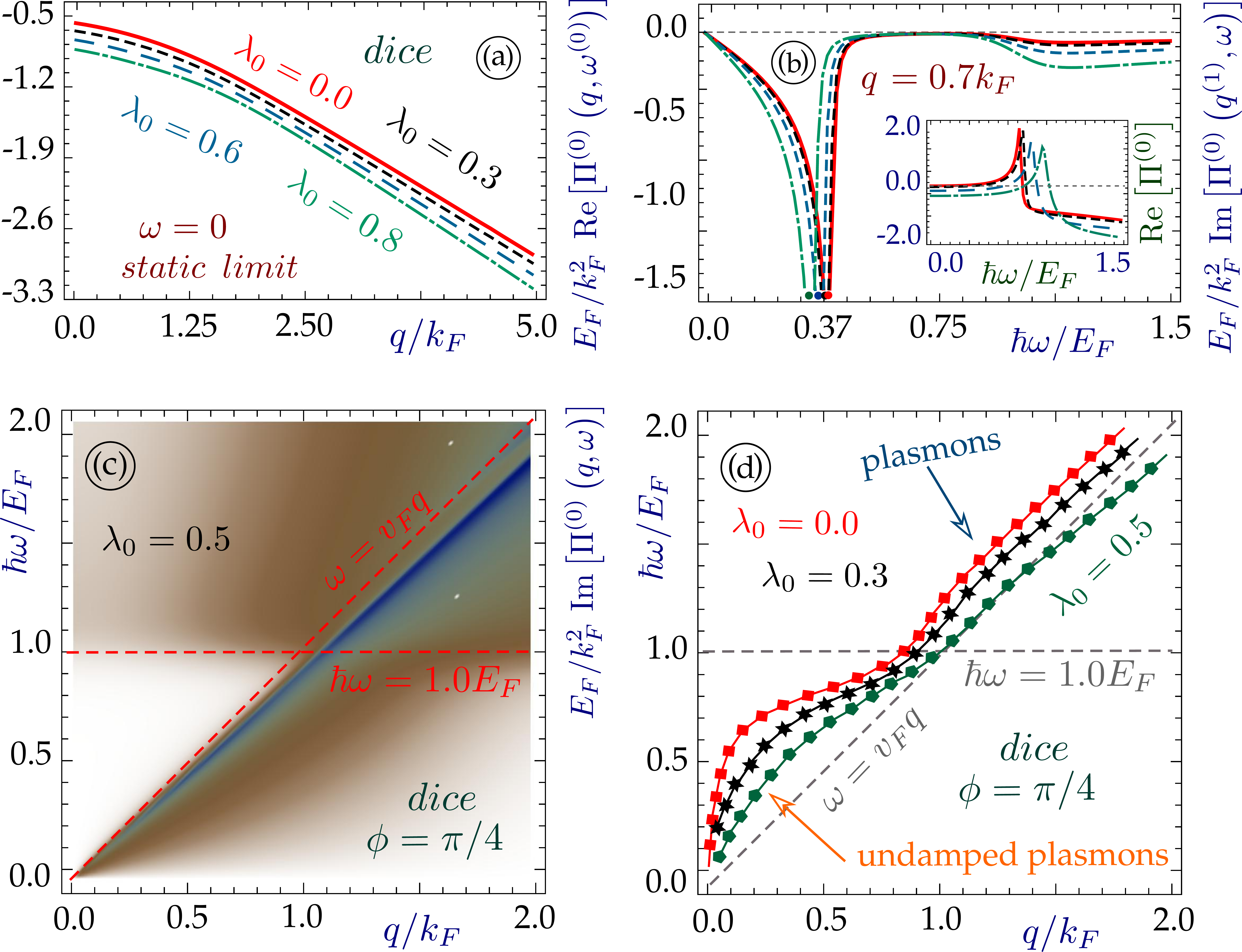} 
\caption{(Color online) $\Pi_0(q, \omega\, \vert \, \lambda_0)$ in units of $k_F^2/E_F$ [$(a)$-$(c)$] 
and plasmon dispersions $(d)$ for a dice lattice ($\phi=\pi/4$) under a circularly-polarized laser field with various $\lambda_0$ values. 
Panel $(a)$ shows the $q$ dependence of $\text{Re}\,[\Pi_0(q, \omega=0 \, \vert \, \lambda_0)]$ in the static limit $\omega = 0$. Plot $(b)$ presents
the $\hbar\omega$ dependence of $\text{Im}\,[\Pi_0(q, \omega \, \vert \, \lambda_0)]$ at $q/k_F=0.7$, while its inset displays the $\hbar\omega$ dependence for $\text{Re}\,[\Pi_0(q, \omega \, \vert \, \lambda_0)]$. 
Plot $(c)$ demonstrates the particle-hole modes $\text{Im}\,[\Pi_0(q, \omega \, \vert \, \phi,\lambda_0)]\neq 0$ at $\lambda_0 = 0.5$.   
}
\label{FIG:3} 
\end{figure}

\medskip
\par
The above electron dressed states share a similarity with an irradiated dice material but are not equivalent to those from 
a gapped Hamiltonian with an added $\Delta_{0}\hat{\Sigma}^{(3)}_{z}$ term, where $\hat{\Sigma}^{(3)}_{z}$ represents a ($3\times 3$) $z$-Pauli matrix with 
the main diagonal $\{1,0,-1\}$ and isused to describe the effect of a point defect.\,\cite{gus19} 

\section{Plasmon mode and optical current}
\label{sec-3}

The self-sustaining charge-density longitudinal oscillations, i.e., plasmons,  play an important role in determining the optical-current properties of low-dimensional
structures.\,\cite{po1,wun,pavlo,SilMain,sch} These include exotic fullerenes and spherical graphitic particles.\,\cite{ourbuck1,b3,ourbuck2}

\medskip
\par
The dispersion relation of plasmon modes in the wavevector-frequency $(q,\omega)$-plane is generally determined from the zero of a dielectric function
$\epsilon(q,\omega\,\vert\,\phi,\lambda_0)$. In terms of the dynamical polarization function $\Pi_0(q,\omega\,\vert\,\phi,\lambda_0)$, we can write  $\epsilon(q,\omega\,\vert\,\phi,\lambda_0)=1-(2\pi\alpha_0/q)\,\Pi_0(q,\omega\,\vert\,\phi,\lambda_0)$, 
where $\alpha_0=e^2/(4\pi\epsilon_0\epsilon_r)$, $\epsilon_r$ is the host-material dielectric constant, and the summation over the valley index $\tau$ is performed. 
In addition, we also require $\Pi_0(q,\omega\,\vert\,\phi,\lambda_0)$ in calculating screening to impurity scattering for electron conduction current. 
The screened potential for a dilute distribution of impurities embedded in a dice lattice  has been 
discussed\,\cite{malcolmMain} as well as for general $\alpha-\mc{T}_3$ materials.\,\cite{ourLor} Furthermore, we notice $\Pi_0(q,\omega\,\vert\,\phi,\lambda_0)$ of
$\alpha-\mc{T}_3$ lattices at $T=0\,$K could substantially differ from that of graphene (red curves for $\phi=0$), as displayed in Fig.\,\ref{FIG:1}. This difference is attributed to
additional channels for electron transitions resulting from the middle flat band, as depicted in the inset of Fig.\,\ref{FIG:1}(b), especially for $\hbar\omega$ close to the 
Fermi energy $E_F$.

\medskip
\par
The complex $\Pi_0(q,\omega\,\vert\,\phi,\lambda_0=0)$ at $T=0\,$K for $0\leq\alpha\leq 1$ (or $0\leq\phi\leq\pi/4$) are presented in Figs.\,\ref{FIG:1}(a)$-$\ref{FIG:1}(d). 
Its imaginary part for $q<k_F$ in Fig.\,\ref{FIG:1}(a) shows a noticeable peak at a lower $\hbar\omega$ in comparison with graphene (red curve). 
For $q>k_F$ in Fig.\,\ref{FIG:1}(b), however, there exists a singular pole scaled as $\backsim - 1/\sqrt{\vert v_F^2q^2 - \omega^2 \vert}$. 
An accompanied zigzag feature in its real part can also be seen in Figs.\,\ref{FIG:1}(c) and \ref{FIG:1}(d), which reveals the $q$-dispersion of a plasmon mode. 

\medskip
\par
Physically, the particle-hole continuum comprising  the single-particle excitation regions are defined as the regions with $\text{Im}\left[\Pi_0(q,\omega\,\vert\,\phi,\lambda_0)\right]\neq 0$.
Once a plasmon branch enters into such region, it will suffer from Landau damping resulting in the decay of a plasmon mode into single-particle
excitations. Thus, we would concentrate on finding the regions of damping-free plasmon modes with $\text{Im}\left[\Pi_0(q,\omega\,\vert\,\phi,\lambda_0)\right] = 0$. 

\medskip
\par
In our calculations, we look for a $(q, \omega)$ region in which plasmon modes could be present. In fact, we find that, for all $\alpha-\mc{T}_3$ materials, 
only one {\it triangle} region appears below the Fermi level $E_F$ plus another one (indicated by a red circle) above the main diagonal ($\omega = v_F q$), as shown in Figs.\,\ref{FIG:2}(a) and \ref{FIG:2}(b). 
Even though there are additional areas free from Landau damping for $q > 2 k_F$,
plasmon modes in free-standing 2D materials cannot exist there. Strictly speaking, a plasmon mode will decay once it goes above the line $\hbar\omega/E_F=1$.
Interestingly, the damping strength varies with $\phi$ and becomes infinitesimally small and even disappears 
for graphene with $\phi \to 0$.

\medskip
\par
The distortion of plasmon dispersions around $\hbar\omega\approx E_F$ in Fig.\,\ref{FIG:2}(c) reflects the contributions associated with electron transitions both starting from and ending in the flat band. 
The influence of this flat band amplifies itself close to the $\hbar\omega/E_F=1$ line, 
where various plasmon branches, corresponding to different $\alpha_0$ values, are expected to be pinched at a single crossing point intersected by the $\hbar\omega/E_F=1$ line and diagonal to $\hbar\omega=\hbar v_Fq$, as found for a dice lattice.\,\cite{malcolmMain} 
Here, however, we find these distorted plasmon branches with $\phi = \pi/6$ are separated from the diagonal $\hbar\omega=\hbar v_F q$ and only display two peaks below and above the $\hbar\omega/E_F=1$ line instead of pinching.
Moreover, various plasmon branches with different $\alpha_0$ values will cross the $\hbar \omega/E_F=1$ line at slightly different $q$ values. 

\medskip
\par
We find one interesting feature by analyzing the degree to which various plasmon modes with a fixed $\alpha_0$ in Fig.\,\ref{FIG:2}(d) are away from the diagonal $\hbar\omega=\hbar v_F q$ 
for the boundary of single-particle excitations. For $\phi\neq 0$, the plasmon energy is always smaller than that of graphene ($\phi = 0$, red curve). 
For all finite $\phi$, there exist two steps for plasmon energies, which are separated by the $\hbar\omega=E_F$ line, except for $\phi = 0$. 
The undamped first step under the $\hbar\omega/E_F=1$ line spans a much larger (up to ten times) $q$ range compared to graphene. 
For $q\ll E_F/(\hbar v_F)$, all plasmon modes become nearly degenerate, corresponding to the electron transitions across the Fermi energy within the upper Dirac cone.
On the other hand, the electron transitions between the flat band and the upper Dirac cone are associated with the range around $q\approx E_F/(\hbar v_F)$ in Fig.\,\ref{FIG:2}(d).
Furthermore, the electron transitions resulting from two Dirac cones relate to the $q\geq 2E_F/(\hbar v_F)$ range. 
In summary, we believe the best condition for observing the $\phi$--dependence of plasmon dispersions and damping is 
around $\phi=0$ because the results for $\phi = \pi/6$ and a dice lattice show only little difference in Fig.\,\ref{FIG:2}(d). 

\medskip 
\par 
Next, we turn our attention to the plasmon dispersions of a dice lattice in the presence of a circularly-polarized laser field. 
The dice lattice ($\phi=\pi/4$) becomes the most different entity compared to graphene ($\phi=0$), for the latter the 
effect of circularly-polarized light is simply adding a band gap (via a $\hat{\Sigma}_z^{(2)}$ term) to the bare Dirac 
Hamiltonian.\,\cite{kiMain,pavlo} The distinctive feature of a dice lattice is that 
$\text{Re}\,[\Pi_0(q,\omega = 0\,\vert\,\lambda_0)]$ no loner becomes a constant within the region of $q < 2\,k_F$, as demonstrated in Fig.\,\ref{FIG:3}$(a)$. Furthermore, as $\lambda_0$ increases from zero, 
the negative peak of $\text{Im}\,[\Pi_0(q,\omega\,\vert\,\lambda_0)]$ shifts downwards in $\hbar\omega$, while the positive peak of $\text{Re}\,[\Pi_0(q,\omega\,\vert\,\lambda_0)]$ shifts upwards, as seen in Fig.\,\ref{FIG:3}$(b)$ and its inset.    

\begin{figure}
\centering
\includegraphics[width=0.6\textwidth]{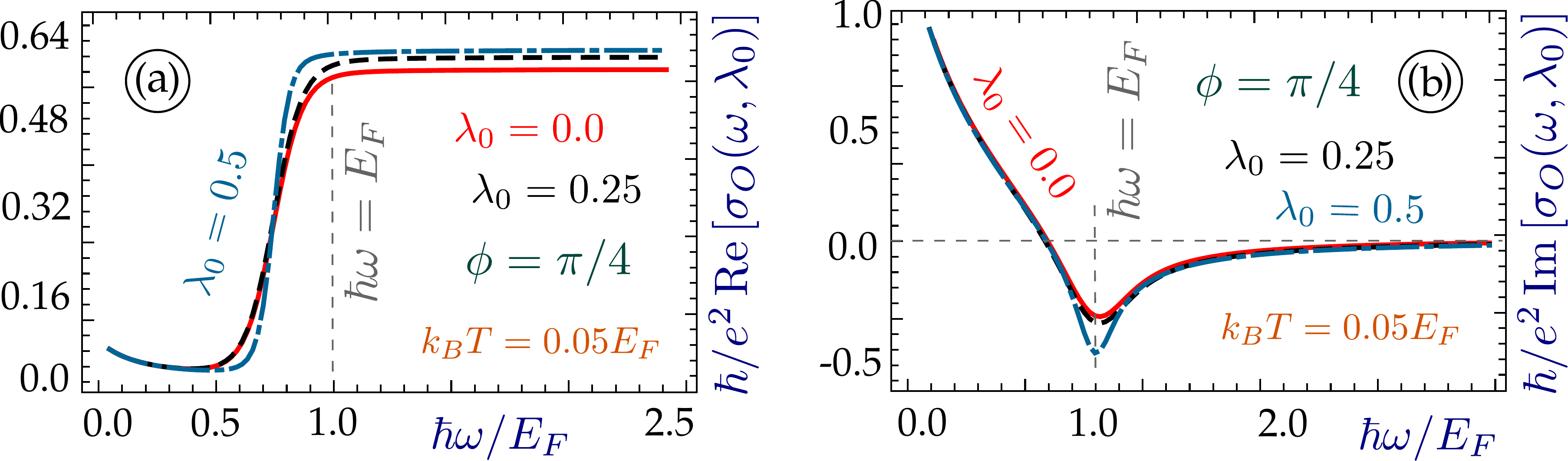}  
\caption{(Color online) Real $(a)$ and imaginary $(b)$ parts of optical-current conductivity $\sigma_O(\omega\,\vert\,\lambda_0)$ in units of $e^2/\hbar$ for an irradiated dice lattice ($\phi=\pi/4$)
as a function of $\hbar\omega$. Here, each curve corresponds to a specific $\lambda_0$ value for fixed $\phi$ and $T$.}
\label{FIG:4}
\end{figure}
 
\medskip
\par
For an irradiated dice lattice, Landau damping is greatly enhanced and shifted downward below the main diagonal $\omega = v_F q$, as found in Fig.\,\ref{FIG:3}(c).  
Although the laser field does not affect the undamped plasmon modes in the triangle region determined by $\omega > v_F q$ and $\hbar \omega/E_F<1$, 
their dispersions are significantly modified beyond this triangle region above the diagonal, as presented in Fig.\,\ref{FIG:3}(d). Meanwhile, the plasmon energy decreases with increasing $\lambda_0$, similar to
the single-electron dispersion in the presence of an energy gap.\,\cite{pavlo} However, no similarity to plasmon dispersions of graphene is found due to the addition of a middle flat band. 
It is worthwhile to mention that the plasmon branch extends into the region below the main diagonal for large $\lambda_0$ values.

\medskip
\par
Finally, we would like to address the issue of laser-induced optical current. The result for the conductivity of a dissipative optical current can be obtained from the calculated 
complex polarization function $\Pi_0(q,\omega \,\vert \, \phi,\lambda_0)$.
Specifically, the optical-current conductivity in the long-wavelength limit is given by 
$\sigma_O(\omega\,\vert\, \phi,\lambda_0)= \lim\limits_{q \to 0}\left\{(ie^2\omega/q^2)\,\Pi_0(q,\omega \,\vert \, \phi,\lambda_0)\right\}$.\,\cite{SilMain} 
Consequently, the real part of $\sigma_O(\omega\,\vert\, \phi,\lambda_0)$ for optical current will correspond to the imaginary part of $\Pi_0(q,\omega \,\vert \, \phi,\lambda_0)$ for absorptive dissipation. 
On the other hand, the imaginary part of $\sigma_O(\omega\,\vert\, \phi,\lambda_0)$ 
will be associated with the real part of $\Pi_0(q,\omega \,\vert \, \phi,\lambda_0)$ for induced polarization. 

\medskip
\par  
Our numerical results for calculated optical-current conductivity in a dice lattice are presented in Fig.\,\ref{FIG:4}. Its real part ${\rm Re}\,[\sigma_O(\omega\,\vert\,\lambda_0)]$ presented in Fig.\,\ref{FIG:4}(a)
reveals that a high plateau starting nearly from $\hbar\omega=E_F$ extends well into a high-frequency region, and it is slightly enhanced by laser irradiation from the result $\backsimeq 1 + 4\lambda^4_0$ 
for graphene.\,\cite{ourcond2018,fp} This is in correspondence with an appearance of Landau damping for plasmon modes in the $q\to 0$ limit, as displayed in Fig.\,\ref{FIG:3}(c).
In addition, for the imaginary part ${\rm Im}\,[\sigma_O(\omega\,\vert\,\lambda_0)]$, we find it independent of $\lambda_0$ except for $\hbar\omega$ close to $E_F$.
Furthermore, a negative peak in ${\rm Im}\,[\sigma_O(\omega\,\vert\,\lambda_0)]$ shows up at $\hbar\omega=E_F$ and becomes sharpened by increasing $\lambda_0$ due to laser irradiation. 
Moreover, ${\rm Im}\,[\sigma_O(\omega\,\vert\,\lambda_0)]$ in the region of $\hbar\omega>2E_F$ is fully suppressed to zero in the long-wavelength limit $q\to 0$.
This is related to the fact that plasmon modes, determined by $\text{Re}\,[\Pi_0(q,\omega\,\vert\,\lambda_0)]=q/2\pi\alpha_0$, do not exist in this region as $q\to 0$, as can be verified from Fig.\,\ref{FIG:3}(d).

\begin{figure}
\centering
\includegraphics[width=0.6\textwidth]{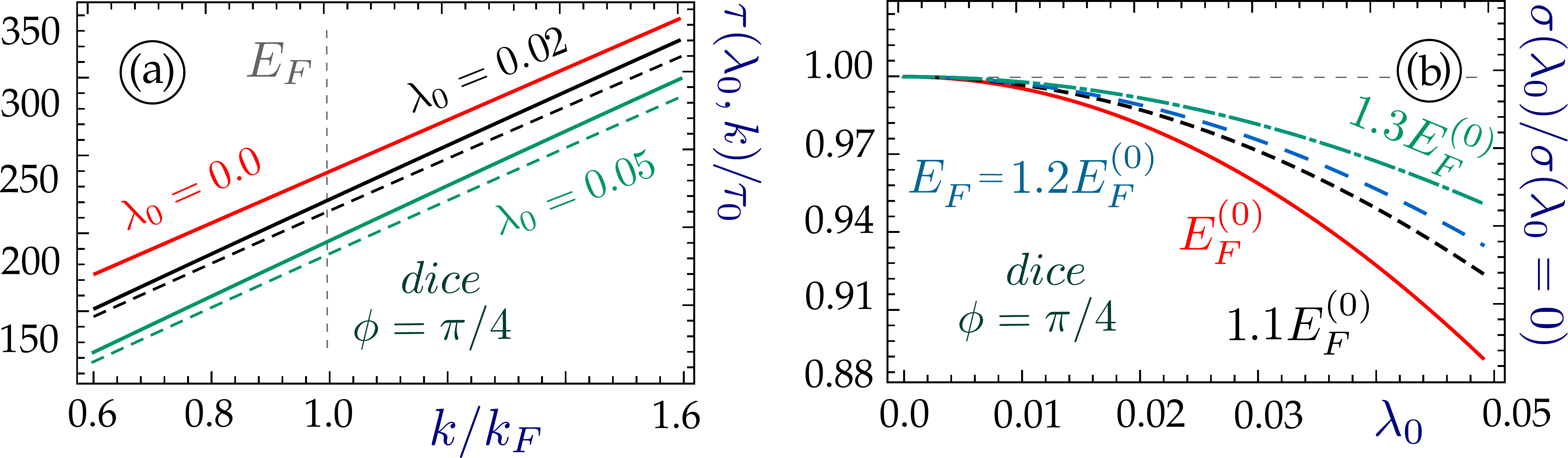} 
\caption{(Color online) Transport conductivity $\sigma_T(\lambda_0)$ for an irradiated dice lattice. Panel $(a)$ displays the relaxation time $\tau(k,\lambda_0)$ in units
of $\tau_0 = 2\hbar E^{(0)}_F/(\pi n_i\alpha_0^2)$ as a function of wavevector $k$ for 
various electron-light coupling constants $\lambda_0$ as labeled, where $E^{(0)}_F$ represents the Fermi energy for the case with ${\cal E}_0=0$. 
The dashed curves show the corresponding results without taking into account the 
laser-induced modification to the static dielectric function. Plot $(b)$ presents the ratio $\sigma_T(\lambda_0)/\sigma_T(\lambda_0 = 0)$ 
as a function of $\lambda_0$ for different Fermi energies.}
\label{FIG:5}
\end{figure}  

\section{Impurity scattering and conduction current}
\label{sec-4}

Now, we consider the transport conductivity $\sigma_T(\lambda_0)$ of an irradiated dice lattice. We will calculate  $\sigma_T(\lambda_0)$ in the relaxation-time
approximation, while the scattering potential is assumed as the point-like Coulomb interaction $U_{im}(r)=Z^*e/ (4\pi\epsilon_0\epsilon_rr)$ with an impurity charge number $Z^*$. 
For finite electron doping $E_F > 0$, the inverse relaxation time is given by\,\cite{asf-rmp, dsmain, castro, ds01} 

\begin{eqnarray}
\label{MainEq}
\nonumber
\frac{1}{\tau(k, \lambda_0)}&=&\frac{n_i}{2\pi \hbar} \, \int\limits_0^{2 \pi} 
 d \beta_{{\bf k},{\bf k'}} \,(1 - \cos \beta_{{\bf k},{\bf k'}}) \\
&\times& \sum\limits_{\tau=\pm 1}\,\int\limits_0^{\infty} \frac{k' dk'}{|\epsilon(\vert \mbox{\boldmath{$k$}} - \mbox{\boldmath{$k$}}' \vert,\omega=0)|^2}\,
\left|\int d^2\mbox{\boldmath{$r$}}\,\Phi_\gamma^\tau(\mbox{\boldmath{$r$}},\mbox{\boldmath{$k$}}',\lambda_0)
\, U_{im}(r) \,\Phi_\gamma^\tau(\mbox{\boldmath{$r$}},\mbox{\boldmath{$k$}},\lambda_0)
\right|^2\delta[\varepsilon_\gamma(k,\lambda_0)-\varepsilon_\gamma(k',\lambda_0)]\ ,
\end{eqnarray}
where $\gamma=+$,
$n_i$ represents the impurity areal density, $\beta_{\,{\bf k},{\bf k'}}$ is the angle between electron wavevectors $\mbox{\boldmath{$k$}}$ and $\mbox{\boldmath{$k$}}'$, 
the full electron wave function is $\Phi_\gamma^\tau(\mbox{\boldmath{$r$}},\mbox{\boldmath{$k$}},\lambda_0)
= \Psi_\gamma^\tau(\mbox{\boldmath{$k$}},\lambda_0) \, \tet{exp}\left(i \, \mbox{\boldmath{$k$}} \cdot \mbox{\boldmath{$r$}} \, \right)$. For isotropic band dispersions and electronic states,
corresponding to circularly-polarized laser irradiation, the relaxation time, $\tau(k, \lambda_0)$, depends only on $k=\vert \mbox{\boldmath{$k$}} \vert$. In the absence of static screening, 
$\tau(k, \lambda_0)$ could be obtained analytically (see Appendix\ \ref{app-3}). 
 
\medskip
\par 
For the isotropic case of circularly-polarized irradiation and at low temperatures, the electric-current $J_0$ per length under an applied DC electric field $E_0$ is expressed as (see Appendix\ \ref{app-3})

\begin{equation}
\label{cur}
J_0 = \left( \frac{e}{\pi} \right)^2E_0 \, \int d^2\mbox{\boldmath{$k$}}\,\left[v_\gamma(k,\lambda_0) \right]^2  \tau(k,\lambda_0)  \, \delta[\varepsilon_\gamma(k,\lambda_0) - E_F] \ ,
\end{equation}
where $v_\gamma(k,\lambda_0)=(1/\hbar)\,\partial\varepsilon_\gamma(k,\lambda_0)/\partial k$ is the electron group velocity.
 
\medskip
\par 
The inverse relaxation time $1/\tau_0(k)$ for a non-irradiated dice lattice is $3/4$ times of that of graphene due to the change of wave-function overlap factors, 
which for graphene is equal to $(1 + \cos{\beta_{\,{\bf k}, {\bf k}'}})^2/4$. Dice lattice and graphene 
represent two limiting cases of $\alpha-\mc{T}_3$ lattices, and their $\tau_0(k)$ has already been calculated.\,\cite{ourLor} Once the circularly-polarized laser is applied, 
the ratio of inverse relaxation times becomes $\tau_0(k)/\tau(k,\lambda_0)\backsimeq 3 \pi/4 - 7 \pi/16 \, \xi^2 + \cdots$, where $\xi 
= c_0 \lambda_0/E_F = (e \mc{E}_0/\hbar\omega)^2(v_F/\omega k_F)$. This apparently leads to a substantial drop of $\sigma_T(\lambda_0)$ in the presence of 
a laser field ${\cal E}_0$. The exact ratio is predetermined by a pseudospin-1 wave function with three inequivalent components, and is not valid for 
graphene or a $2D$ electron gas.
 
\medskip
\par 
Another factor for laser-induced reduction of $\sigma_T(\lambda_0)$ comes from the decreased 
electron group velocity $v_\gamma(k,\lambda_0)$. From Eq.\,\eqref{zero}, we find $[v_F(\lambda_0)]^2\backsimeq v_F^2 
\left(1 - \lambda_0^2/4 \right)^2\left[1 -(\lambda_0^2/4)\,(c_0/E_F)^2 \right]$. Here, the two terms are related to the variation
of the Fermi velocity $v_F$ and the opening of a bandgap $E_G = \lambda_0c_0$, respectively.
Both of these effects lead to a decrease in $\sigma_T(\lambda_0)$ with $\lambda_0$. Using $\mbb{I}_0= 10 \, W/cm^2$ and $c_0/E_F \backsimeq
9.75$, we find the magnitude of the second term is much larger than the first term. 
Besides, the screening factor is also reduced in the presence of a laser field which must be taken into account for an accurate determination of 
$\sigma_T(\lambda_0)$. 
 
\medskip
\par 
The above obtained results for a dice lattice are quite different from those of graphene. For graphene, we get 
$[v_F(\lambda_0)]^2/v_F^2\backsim 1 - \lambda_0^2 \left( c_0 / E_F \right)^2$, and then, the inverse relaxation time $\tau_0/\tau(k,\lambda_0)\backsim 1 - 
3\lambda_0^2 \left(c_0/E_F\right)^2$,\,\cite{kisrep} as well as the reduced transport conductivity $\sigma_T(\lambda_0)/\sigma_T(\lambda_0=0)\backsim 1 - 4 \lambda_0^2 \left( c_0 / E_F \right)^2$.

\medskip
\par  
Numerical results for the relaxation-time $\tau(k,\lambda_0)/\tau_0$ and the transport-conductivity $\sigma_T(\lambda_0)/\sigma_T(\lambda_0=0)$ in a dice lattice are presented in Fig.\,\ref{FIG:5}, 
including the laser-induced modification to the static screening for elastic scattering between electrons and impurities. 
From Fig.\,\ref{FIG:5}(a), we find the relaxation time $\tau(k,\lambda_0)$ is approximately proportional to $k$ and decreases with increasing $\lambda_0$. Moreover, the 
laser-induced modification (dashed curves) to the static dielectric function is important quantitatively.
From Fig.\,\ref{FIG:5}(b), we know the transport conductivity $\sigma_T(\lambda_0)$ decreases with $\lambda_0$ nonlinearly compared to the result in the absence of a laser field, 
which agrees with our analytical evaluations.
In addition, for fixed $\lambda_0$ this reduction effect becomes less and less significant with increasing Fermi energy since the correction is proportional to $c_0/E_F$.
 
\section{Concluding Remarks and Summary}
\label{sec-5}
 
In this paper, we have calculated and analyzed numerical results for plasmon-mode dispersion and damping as well as their effects on displacement and transport currents 
of electrons in irradiated $\alpha-\mc{T}_3$ materials by a circularly-polarized laser. 
As a result, we conclude that the intensity of a laser field can be used effectively to control both optical and transport conductivities in the system, 
in addition to a tuning of them with a structure parameter $0<\alpha\leq 1$ for different $\alpha-\mc{T}_3$ lattices.
 
\medskip
\par 
In particular, for the whole range of $\alpha$ values, we observe that the tuning of plasmon modes reaches the strongest as $\phi \backsim 0$ on the graphene side, but 
it becomes relatively weak as $\phi \backsim \pi/4$ on the dice-lattice side. 
Meanwhile, a significant increase in the plasmon damping above the Fermi level, accompanied by a change of plasmon dispersion below the Fermi energy, is found with increasing $\alpha$ from zero to one.
Moreover, the pinching of plasmon dispersion around the Fermi energy also shows up, which can be attributed to electron transitions from the middle flat band to the upper Dirac cone.  
 
\medskip
\par 
After a circularly-polarized laser has been applied to a dice lattice, the plasmon mode is modified dramatically by lowering its dispersion curve below the main diagonal.
Meanwhile, the transport conductivity of electrons in a dice lattice decreases with increasing laser intensity. 
These results indicate that electron dynamics under irradiation in graphene is quite different from that in a dice lattice and can be controlled by laser, 
which further implies that such a difference can be tuned by a structure parameter $\alpha$ for $\alpha-\mc{T}_3$ materials.
All of these are expected to provide very useful information and guidance for designing nano-electronic and nano-plasmonic devices based on innovative low-dimensional
$\alpha-\mc{T}_3$ materials. 
 
\begin{acknowledgments}
D.H. would thank the support from the Air Force Office of Scientific Research (AFOSR). D.H is also supported by the DoD Lab-University
Collaborative Initiative (LUCI) program. G.G. would like to acknowledge the support from the Air Force Research Laboratory (AFRL) through 
Grant \#12530960.
\end{acknowledgments}

\newpage 
 
\appendix

\section{Laser-Renormalized Electronic States}
\label{app-1}

The low-energy Hamiltonian for a dice lattice irradiated by a laser field under the off-resonance condition can be obtained from
a perturbation theory by using the Floquet-Magnus expansion\,\cite{ourpeculiar}, given by 

\begin{equation} 
\label{aH}
\mbb{H}_{\tau}(k \, \vert \, \theta_{\bf k}) = - \tau \lambda_0 \, \frac{c_0}{2} \, \hat{\Sigma}_z + \frac{\hbar}{\sqrt{2}} \, \mc{V}_F(\lambda_0) \,
\sum\limits_{s = \pm} \, \hat{\Sigma}_s \, k_\tau^{s} \ ,
\end{equation}
where $c_0 = e \, \mc{E}_0v_F /\omega$ is the interaction energy, $\lambda_0 = c_0/(\hbar \omega)$ is a small dimensionless light-electron coupling constant used for the expansion, 
$\mc{V}_F(\lambda_0) = \left[ 1 - (\lambda_0/2)^2 \, \right]v_F$ is the renormalized Fermi velocity,
$k_{\tau}^{\pm} = \tau k_x \pm i k_y =\tau k \, \tet{e}^{i \tau\theta_{\bf k}}$, and $\theta_{\bf k}=\tan^{-1}(k_y/k_x)$.
Additionally, $\hat{\Sigma}_\pm = \hat{\Sigma}_x \pm i\hat{\Sigma}_y$, where $\hat{\Sigma}_{x,y,z}$
are three-dimensional Pauli matrices defined in Ref.\,[\onlinecite{ourpeculiar}].

\medskip 
\par 
Energy dispersions of the electron dressed states associated with the Hamiltonian in Eq.\,\eqref{aH} are found for $\gamma=\pm$ to be

\begin{eqnarray}
\label{aE}
&& \varepsilon_0(k,\lambda_0) = 0 \hskip0.1in \text{and} 
\\ \nonumber 
&& \varepsilon_\gamma(k,\lambda_0) = \gamma \left\{ (\hbar v_F k)^2 \,
\left[ 1 - \left( \frac{\lambda_0}{2} \right)^2 \, \right]^2+c_0^2 \left(\frac{\lambda_0}{2} \right)^2\right\}^{1/2} \ .
\end{eqnarray}
This gives rise to an energy bandgap $E_G= \lambda_0 c_0\equiv 2\Delta_0$ which is exactly one half the graphene bandgap under the same irradiation, 
while renormalized Fermi velocity stays the same as that of graphene.
Clearly, the obtained dispersions do not depend on the valley index $\tau = \pm 1$.

\medskip 
\par 
Furthermore, the electron wave functions for an irradiated dice lattice are calculated as

\begin{equation}
\Psi_\gamma^\tau(\mbox{\boldmath{$k$}},\lambda_0) = \frac{1}{\sqrt{\mc{N}_\gamma^\tau}} \, \left[
\begin{array}{c}
\tau \, \mc{C}^\tau_{1,\gamma}\, \tet{e}^{ - i \tau \theta_{\bf k}} \\
\mc{C}^\tau_{2,\gamma} \\
\tau\,(\hbar v_F k)^2 \, \tet{e}^{ + i \tau \theta_{\bf k}} 
\end{array}
\right] \ ,
\label{wf}
\end{equation}
where

\begin{eqnarray}
&& \mc{C}^\tau_{1,\gamma}(k,\lambda_0) = (\hbar v_F k)^2 + 2 \left(
\delta_\lambda^2 -  \gamma\tau\delta_\lambda\sqrt{
(\hbar v_F k)^2 + \delta_\lambda^2}\,\right) \ , \\
\nonumber 
&&  \mc{C}^\tau_{2,\gamma}(k,\lambda_0) = \sqrt{2} \, \gamma \, (\hbar v_F k) \, \left(
\sqrt{(\hbar v_F k)^2 + \delta_\lambda^2} - \gamma\tau\delta_\lambda\right) \ , \\
\nonumber 
&& \mc{N}^\tau_\gamma(k,\lambda_0 \ll 1) \backsimeq 
4(\hbar v_F k)^4 - 4\gamma \tau\, c_0\lambda_0\, (\hbar v_F k)^3 + 3\left[c_0\lambda_0 \,(\hbar v_F k)\right]^2 + ...\ .
\end{eqnarray}
Here, our parameter $\delta_\lambda = 2 \lambda_0 \, c_0 / (4 - \lambda_0^2)$ is different from the
energy gap $E_G=\lambda_0 c_0=2\Delta_0$. For $\gamma=+1$,
the wave function in Eq.\,\eqref{wf} for the conduction band,
can be simply rewritten as 

\begin{equation}
\Psi_\gamma^\tau(\mbox{\boldmath{$k$}},\lambda_0)=  \left[
 \begin{array}{c}
  \tau \, c_1^{(+1)} \, \tet{e}^{ - i \tau \theta_{\bf k}} \\
  c_2^{(+1)} \\
  \tau \, c^{(+1)}_3\,\tet{e}^{ + i \tau \theta_{\bf k}} 
 \end{array}
 \right] \ ,
\end{equation}
where

\begin{eqnarray}
 && c_1^{(+1)} = \frac{\mc{C}^\tau_{1,\gamma=1}}{\sqrt{\mc{N}^\tau_{\gamma=1}}} \ , 
 \hskip0.2in 
 \left(c_1^{(+1)}\right)^{2} \backsimeq \frac{1}{4} - \frac{\lambda_0}{4} \, \frac{c_0}{\hbar v_F k} \, \tau + 
 \frac{\lambda_0^2}{16} \, \left( \frac{c_0}{\hbar v_F k} \right)^2  ...  \ , 
 \\ \nonumber
 && c_2^{(+1)} = \frac{\mc{C}^\tau_{2,\gamma=1}}{\sqrt{\mc{N}^\tau_{\gamma=1}}} \ ,
 \hskip0.2in 
 \left(c_2^{(+1)}\right)^{2} \backsimeq \frac{1}{2} - \frac{\lambda_0^2}{8} \, \left(\frac{c_0}{\hbar v_F k} \right)^2 + ... \ ,
 \\ \nonumber 
 &&  c_3^{(+1)} = \frac{(\hbar v_F k)^2}{\sqrt{\mc{N}^\tau_{\gamma=1}}} \ ,
 \hskip0.2in 
 \left(c_3^{(+1)}\right)^{2} \backsimeq \frac{1}{4} + \frac{\lambda_0}{4} \, \frac{c_0}{\hbar v_F k} \, \tau + 
 \frac{\lambda_0^2}{16} \, \left( \frac{c_0}{\hbar v_F k} \right)^2 + ...  \ . 
\end{eqnarray}

\par 
\medskip 
For the flat band, on the other hand, we obtain

\begin{equation}
 \label{psi0}
\Psi_0^\tau(\mbox{\boldmath{$k$}},\lambda_0) = \frac{1}{\sqrt{\mc{N}^\tau_{\gamma=0}}} \, \left[
 \begin{array}{c}
 \hbar v_F k \, \tet{e}^{ - i \tau \theta_{\bf k}} \\
 2 \sqrt{2} \, c_0\lambda_0/(4 - \lambda_0^2) \\
 - \hbar v_F k \, \tet{e}^{ + i \tau \theta_{\bf k}} 
 \end{array}
 \right] \ ,
\end{equation}
where

\begin{equation}
\mc{N}^\tau_{\gamma=0}(k,\lambda_0 \ll 1) \backsimeq 2  
( \hbar v_F k)^2 + \frac{1}{2} \, \left(\lambda_0c_0 \right)^2    + ... \ . 
\end{equation}
Here, the wave-function components are no longer equal to each other, as expected for a finite energy gap.

\par 
\medskip 
The obtained wave function\,\eqref{psi0} could be rewritten as 

\begin{equation}
\Psi_0^\tau(\mbox{\boldmath{$k$}},\lambda_0) =  \left[
\begin{array}{c}
c_1^{(0)} \, \tet{e}^{ - i \tau \theta_{\bf k}} \\
c_2^{(0)} \\
- c_1^{(0)} \, \tet{e}^{ + i \tau \theta_{\bf k}} 
\end{array}
\right] \, ,
\end{equation}
where 

\begin{eqnarray}
 && c_1^{(0)} = \frac{\hbar v_F k}{\sqrt{\mc{N}_0}} \, , 
 \hskip0.2in 
 \left(c_1^{(0)}\right)^{2} \backsimeq \frac{1}{2} - \frac{1}{8} \, \left(\frac{\lambda_0 c_0}{\hbar v_F k} \right)^2 + ...  \ , 
 \\ \nonumber
 && c_2^{(0)} = \frac{2 \sqrt{2} \lambda_0}{4 - \lambda_0^2} \, \frac{c_0}{\sqrt{
 \mc{N}_0}
 } \ , 
 \hskip0.2in 
  \left(c_2^{(0)}\right)^{2} \backsimeq \frac{1}{4} \, \left(\frac{\lambda_0 c_0}{\hbar v_F k} \right)^2 + ... \ .
\end{eqnarray}
Here, it is important to notice that the laser-induced corrections to the flat band wave function do not depend on the valley index $\tau$ in contrast to the cases with $\gamma=\pm 1$. 

\section{Wave-Function Overlap}
\label{app-2}

The prefactor, or the overlap of two electronic states, is defined by a scalar product   
$\mbb{O}^\tau_{\gamma, \gamma'}(\mbox{\boldmath{$k$}},\mbox{\boldmath{$k$}}'\,\vert\,\phi,\lambda_0)$ of the initial  
$\Psi_{\gamma}^\tau(\mbox{\boldmath{$k$}},\lambda_0)$
and scattered 
$\Psi_{\gamma'}^\tau(\mbox{\boldmath{$k$}}',\lambda_0)$
electronic states with the wave vectors $\mbox{\boldmath{$k$}}$ and $\mbox{\boldmath{$k$}}'= \mbox{\boldmath{$k$}} + \mbox{\boldmath{$q$}}$

\begin{eqnarray}
\label{SO}
 &&  \mbb{O}^\tau_{\gamma, \gamma'}(\mbox{\boldmath{$k$}},\mbox{\boldmath{$k$}} + \mbox{\boldmath{$q$}}\,\vert\,\phi,\lambda_0) = 
 \Big| \mbb{S}^\tau_{\gamma,\gamma'} (\mbox{\boldmath{$k$}},\mbox{\boldmath{$k$}} + \mbox{\boldmath{$q$}}\,\vert\,\phi,\lambda_0) \Big|^2 \ , \\
 \nonumber
 && \mbb{S}^\tau_{\gamma,\gamma'} (\mbox{\boldmath{$k$}},\mbox{\boldmath{$k$}} + \mbox{\boldmath{$q$}}\,\vert\,\phi,\lambda_0) = 
 \left\langle \,
 \Psi_{\gamma}^\tau(\mbox{\boldmath{$k$}},\lambda_0)
 \,\Big|\Psi_{\gamma'}^\tau(\mbox{\boldmath{$k$}} + \mbox{\boldmath{$q$}},\lambda_0) \, 
 \, \right\rangle \, , 
\end{eqnarray}
where $k'= \sqrt{k^2 + q^2 + 2 k q \cos\beta_{{\bf k},{ \bf k}'}}$ 
and $\beta_{{ \bf k},{ \bf k}'} = \theta_{{\bf k}'} - \theta_{\bf k}$.

\medskip 
\par
For an irradiated ($\lambda_0 > 0$) dice lattice with $\phi = \pi/4$ and $\tau=+1$, we obtain 

\begin{equation}
 \mbb{S}_{0,+1} (\mbox{\boldmath{$k$}},\mbox{\boldmath{$k$}}'\,\vert\,\phi=\pi/4,\lambda_0)
 = c_1^{(0)}(k) \, c^{(+1)}_1(k') \, \tet{e}^{-i \tau \beta_{{ \bf k},{ \bf k}'}}
 + c_2^{(0)}(k) \, c^{(+1)}_2(k') - c_1^{(0)}(k) \, c^{(+1)}_3(k') \tet{e}^{+ i \tau 
 \beta_{{ \bf k},{ \bf k}'}} \, , 
\end{equation}
which corresponds to the transitions from the flat band $\gamma = 0$ to the conduction band with 
$\gamma' = + 1$, $(0 \leftrightarrow +1$ and back. Similarly, we have 

\begin{equation}
 \mbb{S}_{-1,+1} (\mbox{\boldmath{$k$}},\mbox{\boldmath{$k$}}'\,\vert\,\phi=\pi/4,\lambda_0) =
c_1^{(-1)}(k) \, c_1^{(+1)}(k') \, \tet{e}^{-i \tau \beta_{{ \bf k},{ \bf k}'}} +
c_2^{(-1)}(k) \, c_2^{(+1)}(k') + c_3^{(-1)}(k) \, c_3^{(+1)}(k') \,
\tet{e}^{+i \tau \beta_{{ \bf k},{ \bf k}'}} \, 
\end{equation}
for the transitions between the valence $\gamma = -1$ and conduction band with 
$\gamma' = + 1$, $(-1 \leftrightarrow +1)$, and finally,  

\begin{equation}
\label{oc}
 \mbb{S}_{+1,+1} (\mbox{\boldmath{$k$}},\mbox{\boldmath{$k$}}'\,\vert\,\phi=\pi/4,\lambda_0) =
  c_1^{(+1)}(k) \, c_1^{(+1)}(k') \, \tet{e}^{-i \tau \beta_{{\bf k},{ \bf k}'}} +
c_2^{(+1)}(k) \, c_2^{(+1)}(k') + c_3^{(+1)}(k) \, c_3^{(+1)}(k') \, \tet{e}^{+i \tau 
\beta_{{\bf k},{ \bf k}'}} \ .
\end{equation}
We exclude the remaining possible transitions $-1 \leftrightarrow -1$ inside the valence band and 
between the flat and valences bands $0 \leftrightarrow -1$, which are inactive at zero temperature for
electron doping ($E_F>0$). 
In the absence of irradiation ($\lambda_0=0$), these three overlap factors are given in Table\ \ref{tab1}.

\section{Laser-Renormalized Electron Transport}
\label{app-3}

For a finite electron doping $E_F > 0$, the inverse relaxation time is calculated as
%\begin{widetext}
\begin{eqnarray}
\nonumber
 \frac{1}{\tau(k, \lambda_0)}&=& \frac{n_i}{2 \pi \hbar} \, \int\limits_0^{2 \pi} 
 d \beta_{\,{\bf k},{\bf k'}} \, (1 - \cos \beta_{\,{\bf k},{\bf k'}})\\ 
&\times&\sum\limits_{\tau=\pm 1}\, \int\limits_0^{\infty} \frac{k' dk'}{\left|\epsilon(\mbox{\boldmath{$k$}} - \mbox{\boldmath{$k$}}' \vert,\omega=0)\right|^2} \, 
  \Big|\int d^2\mbox{\boldmath{$r$}}\, \Phi^\tau_{\gamma}(\mbox{\boldmath{$r$}},\mbox{\boldmath{$k$}}',\lambda_0)
  \,U_{im}(r) \,\Phi^\tau_\gamma(\mbox{\boldmath{$r$}},\mbox{\boldmath{$k$}},\lambda_0) 
\Big|^2 \, \delta[\varepsilon_\gamma(k,\lambda_0)-\varepsilon_\gamma(k',\lambda_0)] \ ,\ \ \ \
\label{MainEqA}
\end{eqnarray}
where $\beta_{\,{\bf k},{\bf k'}}$ is the angle between $\mbox{\boldmath{$k$}}$ and $\mbox{\boldmath{$k$}}'$, 
the complete wave function is 
$\Phi^\tau_{\gamma}(\mbox{\boldmath{$r$}},\mbox{\boldmath{$k$}},\lambda_0) = \Psi^\tau_\gamma(\mbox{\boldmath{$r$}},\mbox{\boldmath{$k$}},\lambda_0) \, \tet{exp}\left( i \,\mbox{\boldmath{$k$}} \cdot \mbox{\boldmath{$r$}}\, 
\right)$, and $\gamma=+1$.
For isotropic dispersions and electronic states, corresponding to the circularly-polarized
irradiation, the relaxation time depends only on $k=\vert \mbox{\boldmath{$k$}}\vert$. 

\medskip 
\par
In our analytical evaluation, we neglect the static-screening factor $1/\vert \epsilon(q,\omega=0) \vert^2$ in Eq.\,\eqref{MainEqA}. Since we concentrate on the ratio of two inverse relaxation times with/without irradiation, we expect only the change in static dielectric function with $\lambda_0$ will be ignored. 
We begin with the scattering potential matrix element in Eq.\,\eqref{MainEqA}, given by

\begin{equation}
 \mbb{W}^\tau_\gamma(\mbox{\boldmath{$k$}},\mbox{\boldmath{$k$}}') = \int d^2\mbox{\boldmath{$r$}}\, \Phi^\tau_{\gamma}(\mbox{\boldmath{$r$}},\mbox{\boldmath{$k$}}',\lambda_0)
 \,U_{im}(r) \,\Phi^\tau_\gamma(\mbox{\boldmath{$r$}},\mbox{\boldmath{$k$}},\lambda_0) \ ,
 \end{equation}
where $U_{im}(r) = e^2/ (4 \pi \epsilon_0\epsilon_r r) \equiv \alpha_0/r$. As a result, we get

\begin{equation}
  \mbb{W}^\tau_\gamma(\mbox{\boldmath{$k$}},\mbox{\boldmath{$k$}}') =\alpha_0 \, 
 \mbb{S}^\tau_{\gamma,\gamma} (\mbox{\boldmath{$k$}},\mbox{\boldmath{$k$}}'\,\vert\,\phi,\lambda_0)\,
   \int \frac{d^2\mbox{\boldmath{$r$}}}{r} \, \tet{exp}\left[i(\mbox{\boldmath{$k$}} - \mbox{\boldmath{$k$}}') \cdot \mbox{\boldmath{$r$}} \right]\equiv U_{0}(\vert\mbox{\boldmath{$k$}} - \mbox{\boldmath{$k$}}'\vert)\,\mbb{S}^\tau_{\gamma,\gamma} (\mbox{\boldmath{$k$}},\mbox{\boldmath{$k$}}'\,\vert\,\phi,\lambda_0)\ ,
\end{equation}
where $\mbb{S}^\tau_{\gamma,\gamma} (\mbox{\boldmath{$k$}},\mbox{\boldmath{$k$}}'\,\vert\,\phi,\lambda_0)$ is defined in Eq.\,\eqref{SO} and
 
\begin{equation}
U_{0}(q) =  \alpha_0  \, \int d^2\mbox{\boldmath{$r$}} \, \frac{\tet{exp}\left(i\mbox{\boldmath{$q$}} \cdot\mbox{\boldmath{$r$}} \right)}{r} = 
    \frac{2 \pi \alpha_0 }{q} \, .
\end{equation}

\medskip 
\par
The transition rate in the Born approximation is written as

\begin{equation}
 \mbb{T}^\tau_\gamma(\mbox{\boldmath{$k$}},\mbox{\boldmath{$k$}}') =\frac{2 \pi}{\hbar} \, \Big|  \mbb{W}^\tau_\gamma(\mbox{\boldmath{$k$}},\mbox{\boldmath{$k$}}')
 \Big|^2 \, \delta[\varepsilon_\gamma(k,\lambda_0)-\varepsilon_\gamma(k',\lambda_0)]\ ,
 \label{trans}
\end{equation}
where
\begin{eqnarray}
 && \delta[\varepsilon_\gamma(k,\lambda_0)-\varepsilon_\gamma(k',\lambda_0)] =\frac{\delta(k - k')}{\hbar v_F} \ , \\
 \nonumber 
 && \vert \mbox{\boldmath{$k$}}-\mbox{\boldmath{$k$}}' \vert = 2 k \, \sin \left(\frac{\beta_{{\bf k},{\bf k}'}}{2}\right) \ . 
\end{eqnarray}

\medskip 
\par
By using the result in Eq.\,\eqref{trans}, the inverse relaxation time can be formally written as
 
\begin{equation}
 \frac{1}{\tau_\gamma(k,\lambda_0)} = \sum\limits_{\tau=\pm 1}\,\frac{ n_i}{(2 \pi)^2} \, \int\limits_0^{2 \pi} d\beta_{{\bf k},{\bf k'}} \, (1 - \cos \beta_{{\bf k},{\bf k'}}) 
 \int\limits_0^{\infty} k' dk' \, \mbb{T}^\tau_\gamma(\mbox{\boldmath{$k$}},\mbox{\boldmath{$k$}}')\Big|_{k'=k} \ .
\end{equation}
Specifically,for $\gamma=+1$  we find 

 \begin{eqnarray}
 \label{invt}
 && \frac{1}{\tau(k,\lambda_0)} = \frac{\pi n_i}{2 v_F} \, \left( \frac{\alpha_0}{\hbar} \right)^2 \, \frac{1}{k}\,
 \mbb{I}(k,\lambda_0) \ , \\
  \nonumber 
 && \mbb{I}(k,\lambda_0) = \int\limits_{0}^{2 \pi} d \beta_{{\bf k},{\bf k}'} \, \left(\left\{ 
 \left[ 
 \left(c_1^{(+1)}(k,\lambda_0) \, \right)^2 + \left( c_3^{(+1)}(k,\lambda_0) \, \right)^2 \,
 \right] \cos \beta_{{\bf k},{\bf k}'} + \left( c_2^{(+1)} (k,\lambda_0) \, \right)^2
 \right\}^2\right. \\ \, 
 \nonumber
 &&\left.+  
 \left\{ 
 \left[ 
 \left( c_1^{(+1)}(k,\lambda_0) \, \right)^2  - \left( c_3^{(+1)} (k,\lambda_0) \right)^2  \, \right] 
 \sin\beta_{{\bf k},{\bf k}'}
 \right\}^2\,\right) \ .
 \end{eqnarray}

\medskip 
\par
Finally, by using the relaxation-time approximation, the electric current $\mbox{\boldmath{$J$}}_0$ per length is calculated as

\begin{equation}
\mbox{\boldmath{$J$}}_0 = \left( \frac{e}{\pi} \right)^2 \int d^2\mbox{\boldmath{$k$}}\, \tau(k,\lambda_0)\,\mbox{\boldmath{$v$}}(\mbox{\boldmath{$k$}})\left[\mbox{\boldmath{$E$}}_0\cdot
\mbox{\boldmath{$v$}}(\mbox{\boldmath{$k$}})\right] \, \left[- \frac{\pr f_0[\varepsilon(k,\lambda_0)-\mu_0]}{\pr \varepsilon (k,\lambda_0)}  \right]\ ,
\label{curr}
\end{equation}
where $\mbox{\boldmath{$E$}}_0$ represents the external DC electric field, 
$\mbox{\boldmath{$v$}}(\mbox{\boldmath{$k$}})=(1/\hbar)\,\pr\varepsilon(k,\lambda_0)/\pr\mbox{\boldmath{$k$}}$ is the group velocity of electrons, $f_0[\varepsilon(k,\lambda_0)-\mu_0]=\{1+\exp[(\varepsilon(k,\lambda_0)-\mu_0)/k_BT]\}^{-1}$ 
is the thermal-equilibrium distribution function for electrons, $\mu_0$ is the chemical potential, 
and $T$ is the system temperature.
If $T=0\,$K, we simply have $\pr f_0[\varepsilon(k,\lambda_0)-\mu_0]/\pr \varepsilon (k,\lambda_0)=
\delta[\varepsilon(k,\lambda_0) - E_F]$ with Fermi energy $E_F$ and the integral in Eq.\,\eqref{curr} can be performed analytically. 

\begin{table}
	\label{tab1}
	\begin{center}
		\begin{tabular}{c|c|c}
			\hline 
			\hline 
			Material  \hskip0.2in & \hskip0.2in Overlap \hskip0.2in & \hskip0.2in Inverse relaxation time  \\
			\hline 
			Graphene \hskip0.2in & \hskip0.2in $(1 + \cos \beta_{{ \bf k},{ \bf k}'})/2$ \hskip0.2in & \hskip0.2in 
			$\pi$ \\
			\hline 
			Dice lattice \hskip0.2in & \hskip0.2in $(1+ \cos \beta_{{ \bf k},{ \bf k}'})^2/4$  \hskip0.2in & \hskip0.2in 
			$3 \pi/4$ \\
			\hline 
			$\alpha-\mc{T}_3$ \hskip0.2in & \hskip0.2in $(1/4)\left[ \left(1 + \cos \beta_{{ \bf k},{ \bf k}'} \right)^2 + 
			\cos^2 (2 \phi) \sin^2 \beta_{{ \bf k},{ \bf k}'} \right]$ \hskip0.2in & \hskip0.2in 
			$(\pi/8) \, [7 + \cos (4 \phi)]$ \\
			\hline 
			\hline 
		\end{tabular}
	\end{center}
	\caption{Prefactors (wave-function overlaps), and inverse relaxation time $\mbb{I}(k)$ factor from 
		Eq.\,\eqref{invt} for graphene and
		general $\alpha-\mc{T}_3$ materials in the absence of external irradiation ($\lambda_0=0$).}
\end{table}

%\nocite{*}
\bibliography{TPE-R}% Produces the bibliography via BibTeX.

\end{document}